\documentclass[12pt]{book}
     \makeatletter
     \def\nothing#1{}
     \newdimen\earraycolsep
     \def\eqnarray{\let\@currentlabel\theequation
     \global\@eqnswtrue\m@th
     \global\@eqcnt\z@\tabskip\@centering\let \\\@eqncr
     $$\halign to\displaywidth\bgroup\@eqnsel\hskip\@centering
     $\displaystyle\tabskip\z@{##}$&\global\@eqcnt\@ne
     \hskip 2\earraycolsep \hfil$\displaystyle{##}$\hfil
     &\global\@eqcnt\tw@ \hskip 2\earraycolsep 
     $\displaystyle\tabskip\z@{##}$\hfil
     \tabskip\@centering&\llap{##}\tabskip\z@\cr}
     \renewcommand{\theequation}{\arabic{equation}}
     \renewcommand{\thetable}{\arabic{table}}
     \renewcommand{\thefigure}{\arabic{figure}}
     \def\title{\chapter}
     \renewcommand\chapter{\ifodd\c@page\clearpage\else\cleardoublepage\fi
		    \global\@topnum\z@
		    \@afterindenttrue
		    \secdef\@chapter\@schapter}
     \def\@makechapterhead#1{%
       \vspace*{120\p@}%
       {\parindent \z@ \raggedright \reset@font
         \bfseries #1\par
         \nobreak
         \vskip 36\p@
       }}
     \def\author#1{{\pretolerance=10000 \raggedright \advance \leftskip by 
          1in \noindent #1 \vskip 1pc}}
     \def\affiliation#1{{\advance\leftskip by 1in \noindent #1 \vskip -1pc}}
     \def\refnote#1{{$^{\hbox{\scriptsize #1}}$}}
     \def\affnote#1{\llap{$^{\hbox{\scriptsize #1}}$}}
     \def\tablenote#1{\setbox0=\hbox{$^{\hbox{\scriptsize #1}}$}
         \noindent\hangindent=\wd0 \box0}
     \renewcommand\section{\@startsection{section}{1}{\z@}{2pc \@plus 1ex minus
         .2ex}{1pc \@plus .2ex}{\reset@font\normalsize\bfseries}}
     \renewcommand\subsection{\@startsection{subsection}{2}{\z@}{1pc \@plus 1ex
         minus.2ex}{1pc \@plus .2ex}{\reset@font\normalsize\bfseries}}
     \renewcommand\subsubsection{\@startsection{subsubsection}{3}{\parindent}
     	{1pc \@plus 1ex minus.2ex}{-0.5em}{\reset@font\normalsize\bfseries}}
     
     \def\AmS{{\protect\the\textfont2 
         A\kern-.1667em\lower.5ex\hbox{M}\kern-.125emS}}

     \def\p@LaTeX{{\family{times}\series{m}\shape{n}\selectfont 
         L\kern-.36em\raise.3ex\hbox{\scriptsize A}\kern-.15em 
         T\kern-.1667em\lower.7ex\hbox{E}\kern-.125emX}}
     
     \newlength{\colwidth}
     
     \def\@oddhead{\hfil}
     \def\@evenhead{\hfil}
     \def\@oddfoot{{\bfseries\hfil\thepage}}
     \def\@evenfoot{{\bfseries\thepage\hfil}}
     \def\fnum@figure{\footnotesize\raggedright{\bfseries 
          \figurename~\thefigure.}}
     \def\fnum@table{\normalsize\raggedright{\bfseries \tablename~\thetable.}}
     \long\def\@makecaption#1#2{\vskip 10\p@ {#1 #2\par}}
     \long\def\@makefntext#1{\setbox0=\hbox{$\m@th^{\@thefnmark}$}
          \noindent\hangindent=\wd0 \box0 #1}
     \def\centerfig#1#2#3#4{\vspace*{#2}\relax
         \centerline{\hbox to#1{\special{#4:#3.#4 x=#1, y=#2}\hfil}}}
     \newbox\@atbox
     \long\def\atable#1#2#3{\begin{table}[tbp]\centering\footnotesize
     \setbox\@atbox\hbox{#2}
     \parbox{\wd\@atbox}{\caption{#1}}\par\smallskip
     #2
     \par\smallskip\parbox{\wd\@atbox}{\raggedright #3}
     \end{table}}
     \def\@bibitem{\noindent \hangindent=2pc \hangafter=1}
     \def\thebibliography{%
     \section*{REFERENCES}%
     \bgroup\footnotesize
     \def\newblock{\hskip .11em plus.33em minus.07em}%
     \let\bibitem\@bibitem}
     \def\endthebibliography{\par\egroup}
     \def\@nbibitem#1{\noindent \hangindent=2pc \hangafter=1
     \refstepcounter{enumi}\hbox to 2pc{\arabic{enumi}.\hfil}%
     \immediate\write\@auxout{\string\bibcite{#1}{\arabic{enumi}}}}
     \def\numbibliography{%
     \section*{REFERENCES}%
     \bgroup\footnotesize
     \setcounter{enumi}{0}%
     \def\newblock{\hskip .11em plus.33em minus.07em}%
     \let\bibitem\@nbibitem}
     \def\endnumbibliography{\par\egroup}
     \makeatother

\def\fig#1 #2 #3 #4 {\begin{figure} \vspace{#3pt} \caption[#1]{#4} \label{#1} \end{figure}}

\def\figb#1 #2 #3 #4 {\begin{figure}[b] \vspace{#3pt} \caption[#1]{#4} \label{#1} \end{figure}}

\def\figpb#1 #2 #3 #4 {\begin{figure}[b] \vspace{#3pt} \caption[#1]{#4} \label{#1} \end{figure}}

\def\ess{\hskip.444444em plus .499997em minus .037036em}
\def\mss{\hskip.333333em plus .208331em minus .088889em}
\def\sen{\hbox{\scriptsize--}}
\def\eV{e\kern-.10emV }
\def\eVcm{e\kern-.10emV\kern-.15em,\mss}
\def\eVp{e\kern-.10emV\kern-.15em.\ess}
\def\eVc{e\kern-.10emV\kern-.15em/\kern-.10em$c$ }
\def\eVcp{e\kern-.10emV\kern-.15em/\kern-.10em$c$. }

\begin{document}

\setcounter{totalnumber}{2}
\setcounter{topnumber}{1}
\setcounter{bottomnumber}{1}
\renewcommand{\topfraction}{1.0}
\renewcommand{\bottomfraction}{1.0}
\renewcommand{\textfraction}{0.0}

\chapter{REALISTIC EXPANDING SOURCE MODEL FOR RELATIVISTIC HEAVY-ION COLLISIONS}

\author{Scott Chapman\refnote{1} and \underline{J. Rayford Nix}\refnote{1}}

\affiliation{\affnote{1}Theoretical Division\\
Los Alamos National Laboratory\\
Los Alamos, New Mexico 87545}

\section{INTRODUCTION}

An international search is currently underway for the quark-gluon
plasma---a~predicted new phase of nuclear matter where quarks roam almost
freely throughout the medium instead of being confined to individual
nucleons.\refnote{1,2}\ess  Such a plasma could be formed through the
compression and excitation that occur when nuclei collide at relativistic
speeds.  With increasing compression the nucleons overlap sufficiently that
they should lose their individual identity and transform into deconfined
quarks, and with increasing excitation the many pions that are produced overlap
sufficiently that they should lose their individual identity and transform into
deconfined quarks and anti-quarks.

Experimental identification of the quark-gluon plasma, as well as understanding
other aspects of the process, will require knowing the overall spacetime
evolution of the hot, dense hadronic matter that is produced in relativistic
heavy-ion collisions.  The spacetime evolution of this hadronic matter can in
principle be extracted from experimental measurements of invariant one-particle
multiplicity distributions and two-particle correlations in emitted pions,
kaons, and other particles.  The foundations for two-particle correlations were
laid in the 1950s by Hanbury Brown and Twiss,\refnote{3}\mss who used
two-photon correlations to measure the size of stars, and by Goldhaber et
al.,\refnote{4}\mss who used two-pion correlations to measure the size of the
interaction region in antiproton annihilation.  Following this pioneering work,
many researchers have already analyzed correlations among pions and among kaons
produced in relativistic heavy-ion collisions in terms of simple models to
obtain some limited information about the size and duration of the emitting
source.  However, because of the simplicity and/or lack of covariance of the
models that have been used, the spatial and time extensions of the emitting
source resulting from these analyses have frequently been intertangled, and
most of the presently available results may therefore be regarded as
exploratory.

\section{SOURCE MODEL}

We introduce here a new realistic expanding source model for invariant
one-particle multiplicity distributions and two-particle correlations in nearly
central relativistic heavy-ion collisions that contains nine adjustable
parameters, which are necessary and sufficient to properly characterize the
gross properties of the source during its freezeout from a hydrodynamical fluid
into a collection of noninteracting, free-streaming hadrons.  These nine
physically relevant parameters fall into three categories of three parameters
each, with the first category corresponding to the source's longitudinal
motion, the second category corresponding to its transverse motion, and the
third category corresponding to its intrinsic properties.

The three longitudinal parameters are the rapidity $y_{\rm s}$ of the source's
center relative to the laboratory frame (in terms of which the velocity $v_{\rm
s}$ of the source's center relative to the laboratory frame is given by $v_{\rm
s} = \tanh y_{\rm s}$), the longitudinal spacetime rapidity $\eta_0$ of the
right-hand end of the source in its own frame (in terms of which the velocity
$v_{\ell}$ of the right-hand end of the source in its own frame is given by
$v_{\ell} = \tanh \eta_0$), and the longitudinal freezeout proper time
$\tau_{\rm f}$ (in terms of which the longitudinal radius at the end of
freezeout is given by $R_{\ell} = \tau_{\rm f} \sinh \eta_0$).  We assume that
the source is boost invariant within the limited region between its two
ends,\refnote{5,6}\mss and that it starts expanding from an infinitesimally
thin disk at time $t = 0$.

The three transverse parameters are the transverse velocity $v_{\rm t}$ and
transverse radius $R_{\rm t}$ of the source at the beginning of freezeout and a
transverse freezeout coefficient $\alpha_{\rm t}$ that is related to the width
$\Delta\tau$ in proper time during which freezeout occurs and that determines
the shape of the freezeout hypersurface.  The transverse velocity at any point
on the freezeout hypersurface is assumed to be linear in the transverse
coordinate~$\rho$.  As illustrated in Fig.~\ref{hyper} for the reaction
considered here,\refnote{7}$^{\sen}$\refnote{9}\mss
\figb hyper snowbird/hyper/hypero 255 {Freezeout hypersurface, which specifies
the positions in spacetime where the expanding hydrodynamical fluid is
converted into a collection of noninteracting, free-streaming hadrons.}
freezeout proceeds inward from the initial point $\rho = R_{\rm t}$, $z = 0$ to
the source's center and then to the source's ends.  In our model, the
coordinate-space freezeout surface is a hyperboloid of revolution of first one
sheet and then of two sheets, which moves through the expanding source like the
collapsing neck of a fissioning nucleus in the three-quadratic-surface shape
parameterization.\refnote{10}\ess

The three intrinsic parameters are the nuclear temperature $T$, the ratio
$\mu_{\rm b}/T$ of the baryon chemical potential to the temperature (from which
$\mu_{\rm b}$ itself can be readily calculated), and the fraction
$\lambda_{\pi}$ of pions that are produced incoherently.  (The analogous
quantity for kaons is held fixed at unity, in accordance with theoretical
expectations.)

For a particular type of particle, the invariant one-particle multiplicity
distribution and two-particle correlation are calculated in terms of a Wigner
distribution function that includes both a direct term\refnote{11} and a term
corresponding to 10 resonance decays,\refnote{12}\mss namely the decay of meson
resonances with masses below 900 M\eV and strongly decaying baryon resonances
with masses below 1410 M\eVp Integration of the direct part of the Wigner
distribution function over spacetime leads to the Cooper-Frye formula for the
invariant one-particle multiplicity distribution.\refnote{13}\ess

\section{APPLICATION TO NEARLY CENTRAL Si + Au COLLISIONS}

\figpb pip snowbird/pip/piped2 319 {Comparison between model predictions and
experimental data\refnote{7,8} for the invariant $\pi^+$ one-particle
multiplicity distribution.  Results for successively increasing values of the
particle rapidity $y_{\rm p}$ relative to the laboratory frame are multiplied
by 10$^{-1}$ for visual separation.}

As an initial application, we apply our model to the analysis of invariant
$\pi^+$, $\pi^-$, $K^+$, and $K^-$ one-particle multiplicity
distributions\refnote{7,8} and $\pi^+$ and $K^+$ two-particle
correlations\refnote{9} for nearly central Si + Au collisions at $p_{\rm
lab}/A$ = 14.6 G\eVcp  Figure~\ref{pip} shows a comparison between model
predictions and experimental data for an invariant one-particle multiplicity
distribution, and Fig.~\ref{kpcor} shows the dependence of a two-particle
correlation function upon the longitudinal and ``out'' momentum
differences.\refnote{14}\ess

\fig kpcor snowbird/kpcor/kpcor 234 {Dependence of the predicted $K^+$
two-particle correlation function $C$ upon the longitudinal and ``out''
momentum differences, for fixed values of the other three quantities upon which
$C$ depends.}
  
The nine adjustable parameters of our model are determined by minimizing
$\chi^2$ with a total of 1416 data points for the six types of data considered,
so the number of degrees of freedom $\nu$ is 1407.  The error for each point is
calculated as the square root of the sum of the squares of its statistical
error and its systematic error, with a systematic error of 15\% for $\pi^+$,
$\pi^-$, and $K^+$ one-particle multiplicity distributions, 20\% for the $K^-$
one-particle multiplicity distribution, and zero for $\pi^+$ and $K^+$
two-particle correlations.\refnote{7}$^{\sen}$\refnote{9}\ess  The resulting
value of $\chi^2$ is 1484.6, which corresponds to an acceptable value of
$\chi^2/\nu = 1.055$.  The values of the nine parameters determined this way,
along with their uncertainties at 99\% confidence limits on all nine parameters
considered jointly, are given in Table~1.

\atable{Nine adjusted source freezeout parameters.}
{\begin{tabular}{lc}  
\hline \vspace{-10pt} \\

& Value and uncertainty \\
 
Parameter & at 99\% confidence \vspace{2pt} \\
\hline \vspace{-10pt} \\

Source rapidity $y_{\rm s}$ & 1.355 $\pm$ 0.066 \\[2.1pt]

Longitudinal spacetime rapidity $\eta_0$ & 1.47 $\pm$ 0.13 \\[2.1pt]

Longitudinal freezeout proper time $\tau_{\rm f}$ & 8.2 $\pm$ 2.2 fm/\kern-.10em$c$ \\[2.1pt]

Transverse freezeout velocity $v_{\rm t}$ & 0.683 $\pm$ 0.048 $c$ \\[2.1pt]

Transverse freezeout radius $R_{\rm t}$ & 8.0 $\pm$ 1.7 fm \\[1.711pt]

Transverse freezeout coefficient $\alpha_{\rm t}$ & $-0.86$ $^{+0.36}_{-0.14}$ \\[2.1pt]

Nuclear temperature $T$ & 92.9 $\pm$ 4.4 M\eV \\[2.1pt]

Baryon chemical potential $\mu_{\rm b}/T$ & 5.97 $\pm$ 0.56 \\[2.1pt]

Pion incoherence fraction $\lambda_{\pi}$ & \hspace{0pt} 0.65 $\pm$ 0.11 \vspace{2pt} \\
\hline
\end{tabular}} 
  
\atable{Additional calculated freezeout quantities.}
{\begin{tabular}{lc}  
\hline \vspace{-10pt} \\

& Value and uncertainty \\
 
Quantity & at 99\% confidence \vspace{2pt} \\
\hline \vspace{-10pt} \\

Source velocity $v_{\rm s}$ & 0.875 $^{+0.015}_{-0.016}$ $c$  \\[1.711pt]

Longitudinal velocity $v_{\ell}$ & 0.900 $^{+0.023}_{-0.029}$ $c$ \\[1.711pt]

Longitudinal freezeout radius $R_{\ell}$ & 16.9 $^{+5.6}_{-4.9}$ fm \\[1.711pt]

Beginning freezeout time $t_{1}$ & 3.1 $^{+2.5}_{-3.1}$ fm/\kern-.10em$c$ \\[2.1pt]

Freezeout time $t_{2}$ at source center & 8.2 $\pm$ 2.2 fm/\kern-.10em$c$ \\[1.711pt]

Final freezeout time $t_{3}$ & 18.8 $^{+5.8}_{-5.3}$ fm/\kern-.10em$c$ \\[1.711pt]

Freezeout width $\Delta\tau$ in proper time\refnote a & 5.9 $^{+4.4}_{-2.6}$ fm/\kern-.10em$c$ \\[1.711pt]

Baryon chemical potential $\mu_{\rm b}$ & 554 $^{+34}_{-36}$ M\eV \\[1.711pt]

Strangeness chemical potential $\mu_{\rm s}$ & 75 $^{+13}_{-12}$ M\eV \\[1.711pt]

Isospin chemical potential $\mu_{\rm i}$ & $-5.3$ $^{+1.0}_{-1.1}$ M\eV \\[1.711pt]

Number $B_{\rm proj}$ of baryons originating from projectile & 26.1 $^{+8.8}_{-6.6}$ \\[1.711pt]

Number $B_{\rm tar}$ of baryons originating from target & 57 $^{+20}_{-15}$ \\[1.711pt]

Total number $B_{\rm tot}$ of baryons in source & 83 $^{+28}_{-21}$ \\[1.711pt]

Baryon density $n_1$ at beginning of freezeout\refnote b & 0.057
$^{+\infty}_{-0.032}$ fm$^{-3}$ \\[1.711pt]

Baryon density $n_{\rm s}$ along symmetry axis & 0.0222 $^{+0.0097}_{-0.0069}$ fm$^{-3}$ \vspace{2pt} \\
\hline
\end{tabular}} 
{\tablenote{a}Calculated under the additional assumption that the exterior
matter at $z = 0$ that freezes out first has been moving with constant
transverse velocity $v_{\rm t}$ from time $t = 0$ until time $t_1$.

\tablenote{b}The upper limit of $\infty$ for this quantity arises because the
beginning freezeout time $t_1$ could be zero, at which time the shape is an
infinitesimally thin disk.}
  
From these underlying nine adjustable parameters we are able to calculate
several additional freezeout quantities of physical interest and their
uncertainties at 99\% confidence limits, as shown in Table~2.  For this
reaction, the source-frame longitudinal velocity $v_{\ell}$ is 0.900
$^{+0.023}_{-0.029}$ $c$ and the transverse freezeout velocity $v_{\rm t}$ is
0.683 $\pm$ 0.048 $c$.  These relatively large values mean that the
contribution to the energy of the produced pions and kaons from collective flow
is substantial, which leads to a moderately low freezeout nuclear temperature
$T$ of 92.9 $\pm$ 4.4 M\eVp   The transverse radius $R_{\rm t}$ at the
beginning of freezeout is 8.0 $\pm$ 1.7 fm, and the longitudinal radius
$R_{\ell}$ at the end of freezeout is 16.9 $^{+5.6}_{-4.9}$ fm.

We also investigated the dependence of $\chi^2$ upon nuclear temperature T in
the interval 50 to 250 M\eVcm with the remaining eight parameters varied.  On
both sides of the absolute minimum at $T$ = 92.9 M\eVcm $\chi^2$ rises
steeply.  However, for values of $T$ greater than 129 M\eVcm the minimum
solution switches to an unphysical branch corresponding to the nearly
instantaneous freezeout of a thin disk of large radius.
     
The predictive power of the model can be ascertained by comparing calculated
and experimental quantities that were not included in the adjustment.  For
example, the number $B_{\rm proj}$ of baryons in the source that originate from
the projectile is calculated to be 26.1 $^{+8.8}_{-6.6}$, which agrees well
with the value 28 corresponding to a Si projectile.  In addition, although
proton data were not used in our adjustment, the calculated invariant proton
one-particle multiplicity distribution agrees moderately well with the
experimental distribution for particle rapidities relative to the laboratory
frame of 1.3 or greater.  For lower rapidities, however, the data show far more
low-$p_{\rm t}$ protons than are predicted.  We regard these excess protons as
target spectators that have interacted somewhat, but not enough to be
considered part of the expanding system.

\section{CONCLUSIONS}

We have introduced a new realistic expanding source model containing the
necessary and sufficient number of adjustable parameters for properly
describing invariant one-particle multiplicity distributions and two-particle
correlations in nearly central relativistic heavy-ion collisions.  On the basis
of this model, we have extracted the freezeout properties of the the hot, dense
hadronic matter that is produced in nearly central Si + Au collisions at
$p_{\rm lab}/A$ = 14.6 G\eV by  minimizing $\chi^2$ with a total of 1416 data
points for invariant $\pi^+$, $\pi^-$, $K^+$, and $K^-$ one-particle
multiplicity distributions and $\pi^+$ and $K^+$ two-particle correlations.
The extracted properties include not only the nine adjustable parameters
contained in the model but also several additional freezeout quantities that
can be calculated in terms of the underlying nine parameters.  In all cases, we
have presented uncertainties at 99\% confidence limits.

For the reaction considered here, the extracted freezeout nuclear temperature
$T$ is 92.9 $\pm$ 4.4 M\eVcm the local-rest-frame baryon density $n_{\rm s}$
along the symmetry axis is 0.0222 $^{+0.0097}_{-0.0069}$ fm$^{-3}$, and the
longitudinal freezeout proper time $\tau_{\rm f}$ is 8.2 $\pm$ 2.2
fm/\kern-.10em$c$.  We hope that our model will be used in the future to
systematically analyze other existing and forthcoming data on invariant
one-particle multiplicity distributions and two-particle correlations to
extract freezeout properties as functions of bombarding energy and
target-projectile combinations.  A sharp discontinuity in the dependence of the
extracted freezeout properties upon these quantities could signal the onset of
the long-sought-for quark-gluon plasma or other new physics.

\section{ACKNOWLEDGMENTS}

We are grateful to Arnold J. Sierk for his participation in the early stages of
this work and to T. Vincent A. Cianciolo for permitting us to use his
preliminary data on two-particle correlations in our adjustments.  This work
was supported by the U. S. Department of Energy.

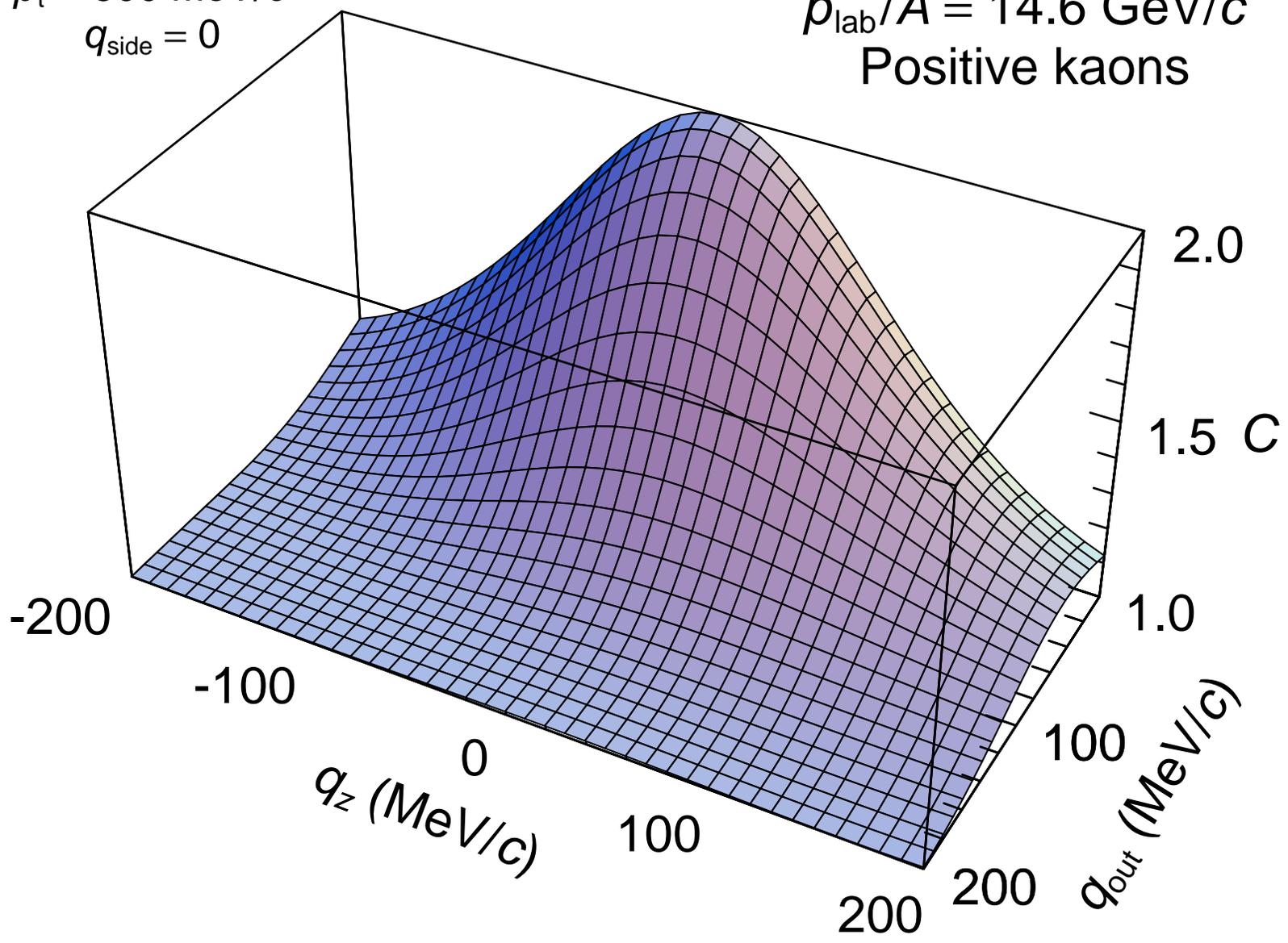
\end{document}